\documentclass[a4paper,11pt]{article}
\usepackage[super,compress]{cite}
\usepackage{amssymb,amsmath,graphicx}
\usepackage{fullpage}

\begin{document}

\begin{center}
\noindent {\bf \Large Analytical solutions for cosmological perturbations in a one-component universe with shear stress}\\
\vspace{5mm}
{Matej \v Skovran\footnote{e-mail address: matej.skovran$@$fmph.uniba.sk}}\\
{\it Department of Theoretical Physics, Comenius University, Bratislava, Slovakia}\\
\end{center}

\begin{abstract}
We construct explicit solutions for scalar, vector and tensor perturbations in a less known setting, a flat universe filled by an isotropic elastic solid with pressure and shear modulus proportional to energy density. The solutions generalize the well known formulas for cosmological perturbations in a universe filled by ideal fluid.
\end{abstract}

\section{Introduction}

In standard cosmological model the anisotropic stress is usually supposed to be zero, or only anisotropic stress coming from viscosity is taken into account. However, the presence of the anisotropic stress significantly modifies the evolution of cosmological perturbations. We study this effect in a model which contains an additional elastic solid continuum and we present analytical solutions of the equations of motion of perturbations in the case of a one-component universe. The results obtained in this way could be helpful for realistic models of the universe to be found in the literature, in which at least a part of present-day dark sector, or the medium driving inflation, is
solid.

The effect naturally depends on the properties of the added continuum, therefore we need a theory describing elastic continuum in the framework of the theory of general relativity at first. Such theory is called relasticity (short name for relativistic elasticity), with Ref. \citenum{CaQu} being the first comprehensive study.

Models based on relastic properties of a solid have already been studied in the
literature. Some of these models extend $\Lambda CDM$ model by adding to it solid matter with negative pressure to density ratio $w$. Cosmic strings ($w=-1/3$) and domain walls ($w=-2/3$) were studied in Ref. \citenum{BuSp} in hope to explain dark matter. The solid was considered there as a candidate for dark energy as well. However, domain walls and cosmic strings are no longer a valid candidate, as follows from the Planck value of the parameter $w$ for dynamical dark energy, $w_{DE} = -1.13^{+0.13}_{-0.10}$ \cite{PLANCK_P}, as well as from the Planck estimates of $w_{DE}$ according to which this parameter is confined to values close to minus one even if it is allowed to vary throughout the evolution of the universe.\cite{PLANCK_P,PLANCK_DE} (Earlier results on the time dependent parameter $w$ can be found in Refs. \citenum{BiWa,BiAlWa,ZhaCriPoZh}.) However, even without microscopic model for the solid, the general framework of the theory of relasticity can be used to describe the dark sector exclusively by means of metric tensor. \cite{BaMo,BaPe}

Another branch of relastic models considers solid matter which acts as an inflation driving medium replacing inflaton field \cite{Gru,EnNiWa,BaMaPeRi,SiSi,Pea,BaMo}.
The first attempts to study cosmological models based on positive pressure solid components ($w>0$) are made in Ref. \citenum{BaSko} and Ref. \citenum{BaSkoCMB}. 

Here we present a set of analytical solutions in the frame of a linearized theory parametrized by a pressure to energy density ratio $w$ and shear modulus to energy density ratio $\xi$. This parametrization is possible due to one component universe simplification only. The simplification is important because the evolution of a scale parameter in time satisfies a power law in a one component universe. The presented solution comprise many models (including models of ideal fluids as a special case $\xi=0$) in the setting of a one dominant component filling a universe.

Some partial results can be found, in different notations, in Refs. \citenum{BaSko,BaSkoCMB}. The discussion of perturbations in a one-component universe in those papers is, however, not too detailed, since the focus is on the examples of two-component universe, with the solid component (radiation-like or stiff) appearing in the universe at some finite time. Also, the analysis is restricted to scalar perturbations there. Here we develop a complete theory of
perturbations in a one-component universe, including solutions for vector and tensor perturbations as well as discussion of singular cases for which the general formulas are not applicable.

\section{Equations of motion}

The dynamics of a spacetime containing relastic continuum is governed by Einstein equations $2G_{\mu\nu}=T_{\mu\nu}$ (we use units $c^4/16\pi G=1$ and $c=1$), which relate Einstein tensor $G_{\mu\nu}$ to the energy-momentum tensor $T_{\mu\nu}$, and equations of state $p_i=p_i(\rho_i)$, which relate the pressure $p_i$ of the $i$-th component of the universe to its energy density $\rho_i$.
We consider a homogeneous and isotropic flat universe (Robertson-Walker metric with $k=0$) with perturbations in the form of planar waves imposed on the metric and energy density. Einstein equations are then reduced to Friedmann equations (zero order equations determining evolution of the scale parameter $a$) and linearized equations describing the evolution of cosmological perturbations.
Friedmann equations are
\begin{equation}\label{eq:Friedmann}
\dot a^2/ a^{2}= \rho/6, \qquad \partial\rho_i/\partial a + 3a^2(\rho_i + p_i)=0,
\end{equation}
where we used total energy density $\rho = \sum \rho_i$. 

The linearized equations governing cosmological perturbations are more complicated and can be found in the textbooks on cosmology, for example in Ref. \citenum{Mu} or Ref. \citenum{Wei}. As a result of linearization, they can be separated into three independent sets of equations, each set describing one type of perturbations: scalar (longitudinal waves), vector (transversal waves) and tensor (gravitational waves).

Usually the continuum is supposed to be an ideal fluid, therefore only the isotropic stress is considered. We extend this standard model by allowing one or more components of the universe to be an isotropic elastic solid. At this point two new parameters $\lambda$ and $\mu$, defined analogically as Lame coefficients in classical elasticity, enter the calculations.
The parameters appear in the expression for the energy density in terms of the contravariant body metric tensor $H^{AB}$ (push-forward of the contravariant spacetime metric tensor to the body space), 
\begin{equation}
\rho = \rho_0 - p \frac{\delta V}{V_0} + \frac{1}{8} \lambda (\Delta H^{~A}_A)^2 + \frac{1}{4} \mu ~\delta H^{~B}_A \delta H^{~A}_B,
\end{equation}
where $\delta V$ is the perturbation to the volume per particle $V \propto a^3$ and the tensor ${\delta H_A}^B$ is the deviation of the actual body metric tensor from the body metric tensor in the partially relaxed state (state with minimum energy $\rho_0$ for given volume $V_0$), with the first index lowered by the covariant body metric tensor in the partially relaxed state. The tensor $\delta {H_A}^B$ can be identified with two times the strain tensor $u_{ij}$ describing deformation of a solid body in nonrelativistic elasticity. The quadratic part of $\rho$ then becomes $(1/2) \lambda u_{ii}^2 + \mu u_{ij}^2$, which is a well-known expression for deformation energy in nonrelativistic elasticity \cite{LaLi}, with Lame coefficients $\lambda$ and $\mu$ characterizing elastic properties of the body. In cosmology, the partially relaxed state is just the state of the matter in an unperturbed universe and the tensor $H^{AB}$ in the comoving gauge reduces to the 3-space part of the contravariant spacetime metric tensor.

Deformations considered here are small perturbations to partially relaxed states. The theory describing them is linear, since the stress tensor is linear in the tensor $\delta {H_A}^B$ characterizing the size of the deformation. A new
element of the theory is the term with the pressure energy density $p$, appearing because we use partially relaxed states instead of completely relaxed one. Given the dependence of $\rho$ on $a$, the $\lambda$ coefficient is redundant as seen from the equation (valid for each component of the universe separately, if they do not interact with each other)
\begin{equation}
\partial p/ \partial a = -a^{-1} (3\lambda + 2\mu)
\end{equation}
This equation is an equivalent of the equation known from non-relativistic elasticity $K=(2/3) \mu + \lambda$, which relates two Lame parameters to the bulk modulus $K= -V dp/dV$. With parameter $\lambda$ being redundant (we use pressure energy density $p$ instead) the remaining parameter describing the properties of the continuum is shear modulus $\mu$. Shear modulus of an ideal fluid is zero.

Other interesting quantities are sound speeds. Let us define an auxiliary variable $c_{s}$ and two physical speeds of sound, $c_{s\parallel}$ (longitudinal) and $c_{s\perp}$ (transversal, see a hint of the derivation in Ref. \citenum{BuSp}),
\begin{equation}
c_{s}^2\equiv \frac{d p}{d\rho} =\frac{\lambda+\frac{2}{3}\mu}{\rho+p}, \qquad c_{s\parallel}^2\equiv \frac{2\mu+\lambda}{\rho+p},\qquad c_{s\perp}^2\equiv \frac{\mu}{\rho+p}.
\end{equation}
The relation $3c_{s}^2=3c_{s\parallel}^2 -4c_{s\perp}^2$ follows immediately from these definitions. These definitions are again valid for each component of the universe separately and characterize sound waves in the respective component. Combined sound speeds (with each variable in the definition summed over the components of the universe) enter the equations for scalar and vector perturbations, since these perturbations are actually sound waves, longitudinal and transversal, in a self-gravitating medium.

In this paper we analytically solve equations of motion for perturbations in a universe filled by a solid continuum with an unperturbed internal geometry ($b_{ij}=0$ in Ref. \citenum{BuSp}), which were derived in Ref. \citenum{PoBa} ($\lambda$ and $\mu$ from this paper denote $(\lambda + \sigma)/V$ and $(\mu + \sigma)/V$ from Ref. \citenum{PoBa} respectively). These equations refer to comoving proper time gauge. However, later we will show connection of the variables characterizing perturbations in this gauge with the more common gauge invariant variables to be found, for example, in Ref. \citenum{Mu}. 

Equations for scalar perturbations from Ref. \citenum{PoBa}, rewritten in a more convenient way (prime denotes differentiation with respect to conformal time $\eta$), are
\begin{equation}\label{eq:scalar}
\frac{a}{a^\prime}s_0^\prime=3c_{s}^2 s_0-3c_{s\parallel}^2s_1,\qquad \frac{a^\prime}{a}s_1^\prime=\frac{1}{3}k^2 s_0 +\frac{1}{4}a^2(\rho+p)\left({s_0-s_1}\right),
\end{equation}
where perturbations $s_0\equiv 6 y_{01} \dot a/a$ and $s_1\equiv y_{11}$ are defined in the appendix. The vector perturbations satisfy 
\begin{equation}\label{eq:vector}
\frac{a}{a^\prime} v_{0\alpha}^\prime=3c_{s}^2
v_{0\alpha}+4c_{s\perp}^2v_{1\alpha},\qquad \frac{a^\prime}{a}v_{1\alpha}^\prime=-\frac{1}{4}k^2 v_{0\alpha}-\frac{1}{4}a^2(\rho + p)\left({v_{0\alpha}- v_{1\alpha}}\right),
\end{equation}
where $v_{1\alpha}\equiv -(a/\dot a) h_{1\alpha}/4$ and $v_{0\alpha}\equiv h_{0\alpha}$ (see the appendix). Index $\alpha$ refers to the transversal direction (for $\vec k=(k,0,0)$ it assumes values $2,3$). Later on we omit this index because the evolution of $v_{0\alpha}, v_{1\alpha}$ is independent of the choice of $\alpha$ (the set of equations for these quantities consists of two independent identical sets, one for each value of $\alpha$).
The tensor perturbations (transversal and traceless part of the 3-metric perturbations denoted as $h_{ij}^T$ in Ref. \citenum{PoBa}) have two degrees of freedom (two polarizations) which both satisfy
\begin{equation}\label{eq:tensor}
h^{\prime\prime}+2\frac{a^\prime}{a}h^\prime+ \left[{k^2+c_{s\perp}^2 a^2(\rho+p)}\right]h=0.
\end{equation}

\section{One-component universe}

The equations governing cosmological perturbations are solvable only numerically in a general multicomponent universe, however analytical solutions do exist in a one-component universe in case the dimensionless parameters $w\equiv p/\rho$ and $\xi=\mu/\rho$ are both constant. The parameter $w$ is a standard parameter used in the description of ideal fluid components of the universe and the parameter $\xi$ quantifies the magnitude of the shear stress of the continuum. 

Friedmann equations are easy to solve in a one-component universe, with the solution $\rho\propto a^{-3(1+w)}$ and $a\propto t^{2/3(1+w)}\propto \eta^{2/(1+3w)}$. Here we constrained the equation of state parameter $w$ to values $w\in \mathbb{R}\backslash \{-1/3,-1\}$ (two excluded values are discussed later). For these values of $w$ the sound speeds are constants and simplify to
\begin{equation}
c_s^2=w, \qquad c_{s\parallel}^2=w+\frac{4\xi}{3(1+w)}, \qquad c_{s\perp}^2=\frac{\xi}{1+w}.
\end{equation}
After all these simplifications, one obtains from the sets of equations for scalar and vector perturbations and the equation for tensor perturbation three second order differential equations for Bessel functions multiplied by an appropriate power of $\eta$:
\begin{equation}\label{eq:s_0(2)}
\eta^2s_0^{\prime\prime}+\frac{4}{1+3w}\eta s_0^\prime +\left[{c_{s\parallel}^2k^2\eta^2+ 4c_{s\perp}^2\frac{6(1+w)}{(1+3w)^2}}\right]s_0=0.
\end{equation}
\begin{equation}\label{eq:v_0(2)}
\eta^2v_0^{\prime\prime}-2\eta v_0^\prime +\left[{c_{s\perp}^2k^2\eta^2+ 3c_{s\parallel}^2\frac{6(1+w)}{(1+3w)^2}}\right]v_0=0.
\end{equation}
\begin{equation}\label{eq:h(2)}
\eta^2 h^{\prime\prime}+\frac{4}{1+3w}\eta h^\prime+\left[{k^2\eta^2+4c_{s\perp}^2 \frac{6(1+w)}{(1+3w)^2}}\right]h=0.
\end{equation}
To express the solutions in a compact way let us define a new parameter $n_\xi$ by the equation 
\begin{equation}\label{eq:n_xi}
n_\xi^2\equiv n_0^2 - 4c_{s\perp}^2\frac{6(1+w)}{(1+3w)^2}, \qquad n_0 \equiv \frac{3(1-w)}{2(1+3w)}
\end{equation}
It will be the order of Bessel functions. Both solutions to the equation are admissible, we can choose for definiteness that with the plus sign. The parameter $n_\xi$ obtained in this way depends on $\xi$ because the sound speed $c_{s\perp}$ does so, and in the special case $\xi=0$ it has the value $n_0$. The solutions of equations (\ref{eq:s_0(2)}), (\ref{eq:v_0(2)}) and (\ref{eq:h(2)}) are:
\begin{equation}\label{eq:s_0_sol}
s_0=\eta^{-n_0}\left[{A_s J_{n_\xi}(\phi_{\parallel})+B_s Y_{n_\xi}(\phi_{\parallel})}\right],
\end{equation}
\begin{equation}\label{eq:v_0_sol}
v_0=\eta^\frac{3}{2} \left[{A_v J_{n_\xi}(\phi_\perp)+B_v Y_{n_\xi}(\phi_\perp)}\right],
\end{equation}
\begin{equation}\label{eq:h_sol}
h=\eta^{-n_0} \left[{A_h J_{n_\xi}(\phi)+B_h Y_{n_\xi}(\phi)}\right].
\end{equation}
The coefficients $A$, $B$ are given by the initial conditions and the arguments of the Bessel functions $J_{n_\xi}$ and $Y_{n_\xi}$ are $\phi_\parallel\equiv c_{s\parallel}k\eta$, $\phi_\perp\equiv c_{s\perp}k\eta$ and $\phi\equiv k\eta$. The solution enters oscillatory mode after the perturbation crosses the respective sound horizon. This behavior of the perturbations, observed also in a universe filled with ideal fluid, follows from the property of Bessel functions that they are maximal approximately when the argument equals their index and for larger arguments they start to oscillate. To complete the picture it remains to write down the functions $s_1$ and $v_1$,
\begin{equation}\label{eq:s_1_sol}
s_1=\eta^{-n_0}\frac{1+3w}{4 c_{s\parallel}^2}\left[{A_s\hat J_{n_\xi}(\phi_{\parallel})+B_s\hat Y_{n_\xi}(\phi_{\parallel})}\right],
\end{equation}
\begin{equation}
8c_{s\perp}^2 v_1=3(1+w)v_0 - \frac{3}{2}(1+3w)\eta^{\frac{3}{2}} \left[{A_s\hat J_{n_\xi}(\phi_{\perp})+B_s\hat Y_{n_\xi}(\phi_{\perp})}\right],
\end{equation}
where we have used auxiliary functions
\begin{equation}
\hat J_{n_\xi}(x)\equiv J_{n_\xi}(x) -\frac{2x}{3}\partial_x J_{n_\xi}(x),\qquad \hat Y_{n_\xi}(x)\equiv Y_{n_\xi}(x) - \frac{2x}{3}\partial_x Y_{n_\xi}(x).
\end{equation}

Asymptotic forms of Bessel functions offer another insight to the behavior of the solutions. The long-wavelength perturbations (perturbations above the sound horizon) are studied in more detail for particular values of $w$ in Ref. \citenum{BaSko} and Ref. \citenum{BaSkoCMB}. 

The solutions offer no direct restrictions on parameters. However, constraints on the parameter $\xi$ can be obtained from physically motivated restrictions to the sound speeds. One constraint follows from the condition that the sound speeds in the continuum should be less than speed of light. If we require the stability of perturbations (real sound speeds) we get another constraint. Put together, the conditions are $0\leq c_{s\parallel}^2\leq 1$ and $0\leq c_{s\perp}^2\leq 1$ which in turn constrain the ratio $\xi/(1+w)\equiv \mu/(\rho + p)$ in the following way
\begin{equation}
-\frac{3}{4}w\leq \frac{\xi}{1+w} \leq \frac{3}{4}(1-w), \qquad 0\leq \frac{\xi}{1+w} \leq 1.
\end{equation}

For example, the radiation-like solid ($w=1/3$) should satisfy $0\leq \xi \leq 2/3$. By arguing that the stability of vector perturbations is not necessary in cosmological setting, because none are produced throughout the inflation, the lower bound could by softened, $-1/3\leq \xi \leq 2/3$ being the bounds.

\section{Special cases}

A special care is required in the case $w=-1$. The parameter $w$ assumes this value throughout the inflation and deeply withing dark energy dominated era as well. For such $w$, the scale parameter grows exponentially, $a\propto e^{Ht}$ ($H$ is the inflationary Hubble parameter, which is constant because energy density stays constant). For $p=-\rho$, the equations (\ref{eq:scalar}) and (\ref{eq:vector}) simplify considerably to
\begin{equation}
s_1=0, \qquad H \dot s_1 = \frac{1}{3}\frac{k^2}{a^2}s_0,\qquad\qquad v_1=0, \qquad H \dot v_1 = -\frac{1}{4}\frac{k^2}{a^2}v_0,
\end{equation}
with trivial solutions $s_0=0$, $s_1=0$ and $v_0=0$, $v_1=0$. The equation governing tensor perturbations stays unchanged and its solution is obtained by performing the limit $w\to -1$ in the general solution (\ref{eq:h_sol}), which can be easily done ($n_\xi=\sqrt{9-24\xi}/2$ and $n_0=-3/2$).

Another special case is $w=-1/3$. It is the case of cosmic strings (since the linear energy density of the string is constant), and also the curvature term satisfies this equation of state. The problem is the choice of conformal time throughout the calculations, because the scale parameter depends exponentially and not polynomially on it. The dependence is $a\propto e^{\mathcal H \eta}=\mathcal H t$, where $\mathcal H\equiv a^\prime/a$ is a constant parameter. Considering this dependence we get simplified sets of equations
\begin{equation}\label{eq:s_w=-1/3}
\frac{1}{\mathcal H}s_0^\prime=\left({1-6\xi}\right)s_1-s_0,\quad \mathcal H s_1^\prime=\frac{1}{3}k^2 s_0 + \mathcal H^2 \left({s_0-s_1}\right),
\end{equation}
\begin{equation}\label{eq:v_w=-1/3}
\frac{1}{\mathcal H}v_0^\prime=6\xi v_1-v_0,\quad \mathcal H v_1^\prime=-\frac{1}{4}k^2 v_0 - \mathcal H^2 \left({v_0-v_1}\right),
\end{equation}
\begin{equation}
h^{\prime\prime}+2\mathcal H h^\prime +\left({k^2 +6\xi \mathcal H^2}\right)h=0.
\end{equation}
The solutions of these equations are no longer combinations of Bessel functions. Instead, they are polynomials when expressed in cosmological time,
\begin{equation}
s_0=A_s t^{m_\parallel-1} + B_s t^{-m_\parallel-1},\quad v_0=A_v t^{m_\perp} + B_v t^{-m_\perp},\quad h=A_h t^{m_\times-1} + B_h t^{-m_\times-1},
\end{equation}
where the three parameters $\mathbf m\equiv (m_\parallel, m_\perp, m_\times)$ are defined in terms of the respective sound speeds $\mathbf c\equiv (c_{s\parallel},c_{s\perp},1)$ as
\begin{equation}
m_p\equiv\left({ 1-c_p^2 k^2\mathcal H^{-2} -6\xi}\right)^\frac{1}{2}.
\end{equation}
The remaining variables $s_1$ and $v_1$ are to be derived from equations (\ref{eq:s_w=-1/3}) of (\ref{eq:v_w=-1/3}) respectively (the calculations are more complicated when $\xi=0$ or $\xi=-1/6$). We do not show these solutions here.

The last special case occurs when $w$ is arbitrary and $\xi=-3w(1+w)/4$ or $\xi=0$. In this case the independent variable $\phi_\parallel$ or $\phi_\perp$ is identically zero ($c_{s\parallel}=0$ or $c_{s\perp}=0$). It is the limit case of general solutions, the respective Bessel functions in the solution stay constant.

\section{Gauge invariant variables}
\label{sec:gauge}

So far we have calculated perturbations in the specific gauge proposed in Ref. \citenum{PoBa}. However, it is often more convenient to work with gauge invariant variables. Tensor perturbations are invariant by themselves, but $s_0$, $s_1$, $v_0$ and $v_1$ are gauge dependent, therefore we must find relations between them and the gauge invariant variables $\Phi$, $\Psi$ and $V_\alpha$ of Ref. \citenum{Mu}.  In this way we replace rather unknown variables by well known gauge invariant variables suitable for further study of perturbations.

Gauge invariant variables also refer to a certain gauge, in particular, the scalar gauge invariant variables are linked to the conformal-Newtonian
gauge.\cite{Mu} Thus, the variables which we have introduced previously and their gauge invariant counterparts to be introduced now describe the same physical object, perturbation to a homogeneous and isotropic universe, in two coordinate systems differing by a small transformation. In the former variables the equations are simpler, while in the latter variables one can use, to a certain extent, Newtonian intuition when interpreting the results (since the variable $\Phi$ is a general relativistic analogue of Newtonian potential).

According to the definitions of the variables $s_0$ and $s_1$ in the appendix (following from definitions of $y_{01}$, $y_{11}$ in Ref. \citenum{PoBa}), the expressions for the potentials $\Phi$ and $\Psi$ are
\begin{equation}\label{eq:PsiPhi}
\Psi=-\frac{\rho + p}{8}\frac{a^2}{k^2} \left({s_0-s_1}\right) ,\qquad \Phi=\Psi -\frac{\mu}{2}\frac{a^2}{k^2} s_1 
\end{equation}
The limit of zero shear stress, for which it holds $\Phi=\Psi$ (standard scenario of a universe filled by an ideal fluid, e.g. in Ref. \citenum{Mu}),  is a good back-check of the calculations.

To parametrize the deviations from $\Lambda$CDM model due to possible modification of the theory of gravity one introduces a set of functions which are equal to one in general relativity, but different from one in modified
theories.\cite{PLANCK_DE} One of these functions, $\tilde \eta = \Phi/\Psi$ (we denote it with a tilde in order to distinguish it from conformal time), is in our model given by 
\begin{equation}
\tilde \eta = 1 + \frac{4\mu}{\rho + p} \frac {s_1}{s_0 - s_1}
\end{equation}
We can see that if we allow for nonzero shear modulus $\mu$, we obtain $\tilde \eta$ different from one even within general relativity. As noted in Ref. \citenum{PLANCK_DE}, there is some tension between the $\Lambda$CDM model and observational data on the
function $\tilde \eta$, especially if a scale-independent parametrization of $\tilde \eta$ is used. Adding a solid component with moderate value of the dimensionless shear modulus $\xi$ to the matter filling the universe might relieve this tension.

Vector perturbations are described by the gauge invariant variable $V_\alpha$, for which we have (see the definition of $h_{1\alpha}$ in the appendix or Ref. \citenum{PoBa})
\begin{equation}
V_{\alpha}=\frac{a}{ik}(\rho+p)v_{0\alpha}.
\end{equation}

After inserting here from equations (\ref{eq:s_0_sol} -\ref{eq:s_1_sol}) we obtain for scalar perturbations:
\begin{eqnarray}
\Psi=\eta^{-n_0-2}\biggl\{{ \phi_{\parallel} \left[{\tilde A_s J_{n_\xi+1}\left({\phi_\parallel}\right)+\tilde B_s Y_{n_\xi+1}\left({\phi_\parallel}\right)}\right]- }\biggr. \nonumber\\ \left.{ -\left[{(n_\xi-n_0)+\frac{8c_{s\perp}^2}{1+3w}}\right] \left[{\tilde A_s J_{n_\xi}\left({\phi_\parallel}\right)+\tilde B_s Y_{n_\xi}\left({\phi_\parallel}\right)}\right] }\right\},
\end{eqnarray}
\begin{eqnarray}
\Phi=\eta^{-n_0-2}\left({1-4c_{s\perp}^2}\right)\biggl\{{  \phi_{\parallel}\left[{\tilde A_s J_{n_\xi+1}\left({\phi_\parallel}\right)+\tilde B_s Y_{n_\xi+1}\left({\phi_\parallel}\right)}\right]- }\biggr. \nonumber\\ \biggl.{ -\left[{(n_\xi-n_0)+ \frac{8c_{s\perp}^2}{1-4c_{s\perp}^2}}\right]\left[{\tilde A_s J_{n_\xi}\left({\phi_\parallel}\right)+\tilde B_s Y_{n_\xi}\left({\phi_\parallel}\right)}\right] }\biggr\},
\end{eqnarray}
for vector perturbations:
\begin{equation}
V =\eta^{-n_0-1}\left[{\tilde A_v J_{n_\xi}(\phi^\xi_\perp)+\tilde B_v Y_{n_\xi}(\phi^\xi_\perp)}\right],
\end{equation}
and equation (\ref{eq:h_sol}) for tensor perturbations. These formulas are valid for $w\in \mathbb{R}\backslash \{-1/3,-1\}$. Again, the sets of coefficients $\tilde A$ and $\tilde B$ are fixed by the initial conditions.

After inserting $\xi = 0$ into equation (\ref{eq:PsiPhi}) we
find that the potentials $\Phi$ and $\Psi$ coincide in the ideal
fluid limit, and we recover the expression for them in a
universe filled with ideal fluid known from the literature
(equation (5.57) in Ref. \citenum{MuFeBra} or equation (7.58) in Ref. \citenum{Mu}),
\begin{equation}
\Phi = \Psi = \eta^{-\nu} \left[{C_1 J_\nu (\sqrt{w}k\eta) + C_2 Y_{\nu}(\sqrt{w}k\eta)}\right], \quad \nu\equiv \frac{5+3w}{2(1+3w)}.
\end{equation}

A qualitatively new property of these solutions is that the oscillatory mode exists even above the horizon. The oscillations are possible because small argument (wavelength above sound horizon) asymptotic formulas of pure imaginary order Bessel functions do oscillate in contrary to the asymptotics of real order Bessel functions. The order of the Bessel functions $n_\xi$ is indeed purely imaginary for sufficiently large shear stress: if we define the critical value of $\xi$ by the equation $n_{\xi_{crit}}=0$, $\xi_{crit}\equiv 3(1-w)^2/32$, the order of the Bessel functions is purely imaginary for $\xi$ above that critical value. 

\section{Conclusion}

We have found explicit solutions for scalar, vector and tensor perturbations in a one-component universe with shear stress. The solutions are combined of Bessel functions and their form in the specific case $\mu=0$ can be easily verified by comparing them with the formulas to be found in the literature \cite{Mu}.

Additionally we pointed out that in the presence of continuum with sufficiently large shear stress in the universe the evolution of the perturbations is modified in a nontrivial way. It turns out that the perturbations oscillate even above the horizon, which is not possible in the standard scenario. 

The results provide a qualitative insight into the topic studied in a simplified setting. The major part of the paper is devoted to a one-component universe. From the physical point of view considering just one kind of matter in the universe is a strong simplification. However, it can yield a good approximation of the evolution of perturbations at times when one component of the universe dominated all the others. Standard cosmology predicts eras throughout which one or another component of the universe is dominant. Our solution can approximate the evolution of cosmological perturbations during such eras.

A solution that is especially significant for standard cosmology is that describing the radiation dominated era ($w=1/3$), which starts after the era of cosmological inflation in standard cosmological scenario. Let us assume that a radiation-like solid was present in the universe during that era, where by radiation-like solid we understand a continuum that satisfies the same equation of state as radiation ($w=1/3$), but has non-zero shear stress. Such continuum  would not change the time dependence of the scale parameter, but the evolution of the cosmological perturbations throughout that era would be modified. Depending on the parameter $\xi$ the perturbations would be enhanced or weakened. When the radiation dominated era ends, the effect of radiation-like solid becomes negligible for further evolution of the cosmological perturbations (as follows from the same considerations as in the standard scenario), but the magnitude of the perturbations stays altered. This offers a possibility to modify the measured magnitudes of primordial perturbations produced throughout inflation, however this goes over the scope of this paper. Let us just mention that the large-scale anisotropies of CMB are strongly affected by the presence of the radiation-like solid, therefore the shear modulus must be very small to be consistent with data \cite{BaSkoCMB}. 

In the previous example, an alternative evolution of cosmological perturbations changes the form of the power spectrum of CMB anisotropies for given cosmological parameters, and in such a way yields different observational values of them. Another possibility is that such evolution leaves the spectrum unchanged but rescales it; that is, it changes its values expressed in terms of the perturbations that had emerged from inflation by a constant factor. This would lead to a modification of parameter space in inflationary scenarios. An example of such effect is given in Ref. \citenum{BaSko}, where the authors consider universe with stiff solid ($w>1/3$) as an additional component.

\appendix

\section{Metric perturbations}
\label{ap}

For the purpose of self-containment, this appendix contains the definitions of metric perturbations used in the paper. Let us define perturbation to the Robertson-Walker metric $\delta g_{\mu\nu}$ in the signature $(+,-,-,-)$ by
\begin{equation}
\delta g_{00}=2\phi,
\end{equation}
\begin{equation}
\delta g_{0i}=a(t)\left({B_{,i}+S_i}\right),
\end{equation}
\begin{equation}
\delta g_{ij}= a^2(t)\left({2\psi\delta_{ij}+2E_{,ij}+F_{i,j}+F_{j,i}+h^{TT}_{ij}}\right).
\end{equation}
Perturbations $\phi$, $B$, $S_i$, $\psi$, $E$, $F_i$ and $h^{TT}_{ij}=h^{TT}_{ji}$ (the notation is in consistency with Ref. \citenum{Mu}, but perturbations are defined in cosmological rather than conformal time) are functions of $x^{\mu}$, satisfying the conditions
\begin{equation}
F_{i,i}=0, \quad S_{i,i}=0,\quad h^{TT}_{ii}=0, \quad h^{TT}_{ij,i}=0.
\end{equation}
The first two conditions say that $F_i$ and $S_i$ are transversal 3-vectors and the next two say that $h^{TT}_{ij}$ is traceless and transversal 3-tensor.

In section \ref{sec:gauge}, two gauge-invariant quantities describing scalar perturbations,
\begin{equation}
\Phi\equiv\phi-{\left[{a\left({B-a\dot E}\right)}\right]}^{\displaystyle .}, \qquad \Psi\equiv\psi+\dot a\left({B-a\dot E}\right),
\end{equation}
are introduced. (The dot denotes differentiation with respect to the cosmological time $t$.) These quantities help us to distinguish physical perturbations from fictitious ones. If both $\Phi$ and $\Psi$ are equal to zero, metric perturbations are fictitious and can be removed by a transformation of coordinates. On the other hand, there is an infinite number of gauge-invariant variables, because any linear combination of $\Phi$ and $\Psi$ is also gauge-invariant. 

For vector perturbations only one gauge-invariant quantity (3-vector)
\begin{equation}\label{eq:V_idef}
V_i\equiv S_i-a\dot F_i
\end{equation}
can be found and tensor perturbations are invariant by definition.

On the other hand, in Ref. \citenum{PoBa} a different approach has been chosen. The comoving proper time gauge is chosen to work in. Just one plane wave, for which the wave covector is chosen to be $(k,0,0)$ (rotational invariance of the background solution allows this without loss of generality of the solution), is introduced in the continuum. As a result, the perturbation to the background metric is 
\begin{equation}
\delta g_{00}=0
\end{equation}
\begin{equation}
\delta g_{0i}\equiv -ikh_{0i}(t)e^{ikx^1},
\end{equation}
\begin{equation}
\delta g_{ij}\equiv a^2(t)h_{ij}(t)e^{ikx^1},
\end{equation}
where the signature is $(-,+,+,+)$. The three modes of cosmological perturbations follow from these definitions.

Tensor perturbations (gravitational waves) are given by a transversal traceless 3-tensor $h^{TT}_{ij}$, the nonzero components of which are $h^{TT}_{23}=h_{23}$ for $\times$ polarization and $h^{TT}_{22}=-h^{TT}_{33}=(h_{22}-h_{33})/2$ for $+$ polarization.

Vector perturbations are described by the vectors $h_{1\alpha}$ and $h_{0\alpha}$, where $\alpha = 2,3$ (two transversal directions). In the paper we have used the variables $v_{1\alpha}\equiv -(a/\dot a)h_{1\alpha}/4 $ and $v_{0\alpha}\equiv h_{0\alpha}$ to describe these perturbations.

The comoving proper time gauge removes the ambiguity in the choice of the coordinates up to the remaining freedom of choosing the zero time hypersurface. Because of this freedom, the three remaining quantities $h_{01}, h_{11}$ and $h_{ii}$ (summed over $i$) describing scalar perturbations can be reduced to two independent quantities $y_{01}$ and $y_{11}$, defined by equations 
\begin{equation}
h_{01}=y_{01}+y,\qquad h_{11}=y_{11}+2\frac{\dot a}{a}y, \qquad h_{ii}=y_{11}+6\frac{\dot a}{a}y,
\end{equation}
where $y$ contains no physical information about scalar perturbations and satisfies the equation $\dot y=-y_{01}(a/\dot a)(\rho + p)/4$. In the paper we have used variables $s_0\equiv 6 y_{01} \dot a/a$ and $s_1\equiv y_{11}$ to describe scalar perturbations.


\end{document}